\documentclass[twocolumn,showpacs,preprintnumbers,amsmath,amssymb,superscriptaddress,aps,prl]{revtex4}

\usepackage{epsfig}
\usepackage{graphicx}
\usepackage{color}

\def\poub#1{}

\def\vert#1{{\color{green}}}
\usepackage{amsmath}

\begin{document}

\title{New determination of the fine structure constant and test of the quantum electrodynamics}

\author{Rym~Bouchendira}
\affiliation{Laboratoire Kastler Brossel, Ecole Normale Sup\'erieure, Universit\'e Pierre et Marie Curie, CNRS, 4 place Jussieu, 75252 Paris Cedex 05, France}
\author{Pierre~Clad\'e}
\affiliation{Laboratoire Kastler Brossel, Ecole Normale Sup\'erieure, Universit\'e Pierre et Marie Curie, CNRS, 4 place Jussieu, 75252 Paris Cedex 05, France}
\author{Sa\"\i da~Guellati-Kh\'elifa}
\affiliation{Conservatoire National des Arts et M\'etiers,
292 rue Saint Martin, 75141 Paris Cedex 03, France}
\author{Fran\c cois~Nez}
\affiliation{Laboratoire Kastler Brossel, Ecole Normale Sup\'erieure, Universit\'e Pierre et Marie Curie, CNRS, 4 place Jussieu, 75252 Paris Cedex 05, France}
\author{Fran\c cois~Biraben}
\affiliation{Laboratoire Kastler Brossel, Ecole Normale Sup\'erieure, Universit\'e Pierre et Marie Curie, CNRS, 4 place Jussieu, 75252 Paris Cedex 05, France}


\begin{abstract}
We report a new measurement of the ratio $h/m_{\mathrm{Rb}}$ between the Planck constant and the mass of $^{87}\mathrm{Rb}$ atom. A new value of the fine structure constant is deduced, $\alpha^{-1}=137.035\,999\,037\,(91)$ with a relative uncertainty of $6.6\times 10^{-10}$. Using this determination, we obtain a theoretical value of the electron anomaly $a_\mathrm{e}=0.001~159~652~181~13(84)$ which is in agreement with the experimental measurement of Gabrielse ($a_\mathrm{e}=0.001~159~652~180~73(28)$). The comparison of these values provides the most stringent test of the QED. Moreover, the precision is large enough to verify for the first time the muonic and hadronic contributions to this anomaly.
\end{abstract}

\maketitle

The fine structure constant $\alpha$ characterizes the strength of the electromagnetic interaction. This dimensionless quantity is defined as:
\begin{equation}
\alpha=\frac{e^2}{4\pi\epsilon_0 \hbar c}, \label{Eq1}
\end{equation}
where $\epsilon_0$ is the permittivity of vacuum, $c$ the speed of light, $e$ the electron charge and
$\hbar $ the reduced Planck constant ($\hbar =h/2\pi$). It appears in the expressions of the ionization energy of hydrogen atom, of the fine and hyperfine structures of atomic energy levels, and it is the parameter of the quantum electrodynamics (QED) calculations. Its measurement in different domains of physics is a test of the consistency of the theory. The most accurate value is deduced from the combination of the  measurement of the electron anomaly $a_\mathrm{e}$ with a very difficult QED calculation. The last result, by Gabrielse at Harvard University, gives a value of $\alpha$ with a relative uncertainty of $3.7 \times 10^{-10}$ \cite{{Gabrielse2008},{Kinoshita}}. Nevertheless this impressive result is fully dependent on QED calculations and can be liable to a possible error: in 2007, Aoyama \emph{et al} detected an error which shifted the $\alpha$ value by $4.7 \times 10^{-9}$ \cite{{Kinoshita},{Gabrielse2006},{Gabrielse2007}}. Consequently, to check these calculations, another determination of $\alpha$ is required. Up to now the other QED  independent determinations of $\alpha$ were less accurate by at least an order of magnitude. The measurement of the quantum Hall effect provides an $\alpha$ value with an uncertainty of $1.8 \times 10^{-8}$ \cite{codata06} and the accuracies of the determinations deduced from the recoil velocity measurement were respectively $7.7 \times 10^{-9}$ and $4.6 \times 10^{-9}$ for the Cesium and Rubidium experiments \cite{{Wicht},{Cadoret}}.

\begin{figure}
\begin{center}
\includegraphics[width = .8\linewidth]{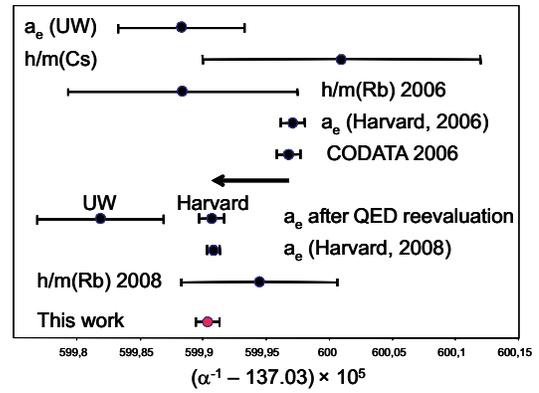}
\end{center}
\caption{Determinations of $\alpha$ with a relative uncertainty smaller than $10^{-8}$; $a_\mathrm{e}$(UW): measurement by Dehmelt at the University of Washington \cite{VanDick}; $h/m_{\mathrm{Cs}}$: measurement of the Cesium recoil velocity at Stanford \cite{Wicht}; $h/m_{\mathrm{Rb}}$: measurement of the Rubidium recoil velocity at Paris in 2006 \cite{Clade2} and 2008 \cite{Cadoret}; $a_\mathrm{e}$(Harvard): measurement of $g-2$ at Harvard University in 2006 \cite{Gabrielse2006} and 2008 \cite{Gabrielse2008}; CODATA 2006: best adjustment by the Committee on Data for Science and Technology \cite{codata06}; the arrow corresponds to the shift of the values of $\alpha$ due to the reevaluation of the QED calculation of $a_\mathrm{e}$ in 2007 \cite{{Gabrielse2007},{Kinoshita}}.}
\label{fig:alpha}
\end{figure}

In this Letter we present a new measurement of the ratio $h/m_{\mathrm{Rb}}$ between the Planck constant and the mass of $^{87}\mathrm{Rb}$ atom and we obtain a new value of $\alpha$:
\begin{equation}
\alpha^{-1}=137.035\,999\,037\, (91). \label{valeuralpha}
\end{equation}
With a relative uncertainty of $6.6\times 10^{-10}$, this value improves our precedent result by a factor of about seven~\cite{Cadoret}. The comparison with the value deduced from the electron anomaly provides the most stringent test of QED \cite{Gabrielse2008}. Indeed there is a very good agreement with this last value ($\alpha^{-1}=137.035\,999\,084\, (51)$) as illustrated on Fig.~\ref{fig:alpha}. This agreement confirms together the recent $g-2$ measurement of Gabrielse by comparison with the value obtained by Dehmelt at the University of Washington \cite{VanDick} and the recent correction found in the calculation of the electron anomaly \cite{Kinoshita}. The discussion on this agreement will be presented at the end of this Letter.

The fine structure constant is deduced from the measurement of $h/m_{\mathrm{Rb}}$ thanks to the relation:
\begin{equation}
       \alpha^2=\frac{2R_\infty}{c}\frac{m_{\mathrm{Rb}}}{m_\mathrm{e}}\frac{h}{m_{\mathrm{Rb}}}, \label{calculalpha}
\end{equation}
where $m_\mathrm{e}$ is the electron mass. In equation (\ref{calculalpha}), the Rydberg constant $R_\infty$ and the mass ratio $m_{\mathrm{Rb}}/m_{\mathrm{e}}$ are known with an accuracy of $7 \times 10^{-12}$ \cite{{codata06},{Udem},{EPJD00}}
and $4.4 \times 10^{-10}$ \cite{{Bradley},{Myers}} respectively: the limiting factor is the ratio $h/m_{\mathrm{Rb}}$. In our experiment, it is deduced from the measurement of the recoil velocity $v_r=\hbar k/m_{\mathrm{Rb}}$ of a Rb atom when it absorbs a photon of momentum $\hbar k$.

The principle of the experiment has been described previously \cite{Cadoret}. It combines a Ramsey-Bord\'e atom interferometer \cite{Borde} with Bloch oscillations (BO). The method is to coherently transfer many recoils to the atoms at rest and to measure the final velocity distribution. The experiment develops in three steps. i) Firstly, a pair of $\pi/2$ pulses of a Raman transition transfers the $^{87}$Rb atoms from the $F=2$ hyperfine sublevel to the $F=1$ one and produces a fringe pattern in the velocity distribution of these atoms. The width of the envelope of this velocity distribution varies inversely with the $\pi/2$ pulse duration $\tau$, while the fringe width varies as $1/T_{\mathrm{R}}$, where $T_{\mathrm{R}}$ is the delay between the two $\pi/2$ pulses. ii) Secondly, we transfer to the selected atoms as many recoils as possible by means of BO. Bloch oscillations have been first observed in atomic physics by the groups of Salomon in Paris and Raizen in Austin \cite{{Dahan},{Peik},{Raizen}}. They can be interpreted as Raman transitions in which the atom begins and ends in the same energy level, so that its internal state $(F=1)$ is unchanged while its velocity has increased by $2 v_r$ per BO. Bloch oscillations are produced in a one dimension vertical optical lattice which is accelerated by linearly sweeping the relative frequencies of the two counter propagating laser beams (frequencies $\nu_1$ and $\nu_2$). This leads to a succession of rapid adiabatic passages between momentum states differing by $2\hbar k$. iii) Finally, the final velocity of the atoms is measured by a second pair of $\pi/2$ pulses which transfers the atoms from the $F=1$ to the $F=2$ hyperfine level. The frequency difference between the two pairs of $\pi/2$ pulses is scanned to obtain a fringe pattern from which the velocity variation between the two pairs of $\pi/2$ pulses is deduced.

The experimental setup uses a double vacuum cell. A two dimensional magneto optical trap (2D-MOT) produces a slow atomic beam (about $10^9$ atoms/s at a velocity of 20 m/s) which loads during 250 ms a three dimensional magneto optical trap. Then a $\sigma^+ - \sigma^-$ molasses generates a cloud of about $2 \times 10^8$ atoms (in the $F=2$ hyperfine level) with a 1.7 mm radius and at a temperature of 4 $\mu$K. A vertical magnetic field of 7 $\mu$T is applied and a radio frequency pulse is used to select atoms in the $F$ = 2, $m_F$ = 0 state: a first pulse transfers the atoms from the $F$ = 2, $m_F$ = 0 Zeeman sub level to the $F$ = 1, $m_F$ = 0 one. A laser beam pushes away the atoms left in the $F$ = 2 hyperfine level and the atoms in the $F$ = 1, $m_F$ = 0 level come back to the $F$ = 2, $m_F$ = 0 level with a second radio frequency pulse. Then we follow the procedure described in reference \cite{Cadoret}. The atoms are accelerated and decelerated by two sequences of 300 BO (duration of 4.6 ms with a delay between them of 10.3 ms) to make an atomic elevator and displace the atomic cloud in the upward or downward direction. The atoms are accelerated in the opposite direction with 500 BO (duration 5.6 ms) before the first pair of $\pi/2$ pulses which produces the Raman transition from the $F$ = 2, $m_F$ = 0 level to the $F$ = 1, $m_F$ = 0 one (delay $T_{\mathrm{R}}$ = 10 ms and duration $\tau$ = 600 $\mu$s). The atoms left in the $F$ = 2 are pushed away. Finally, 500 BO decelerate the atoms before the second pair of $\pi/2$ pulses and the population of the $F$ = 2 and $F$ = 1 levels are measured with a time of flight technique. The laser beams used for the Raman transitions and the Bloch oscillations are sent on the atoms thanks to two optical fibers following the scheme described in reference \cite{Clade2}. They are very well collimated with a beam waist of 3.6 mm and the power of each beam used for the BO is about 150 mW.

\begin{figure}
\begin{center}
\includegraphics[width = .8\linewidth]{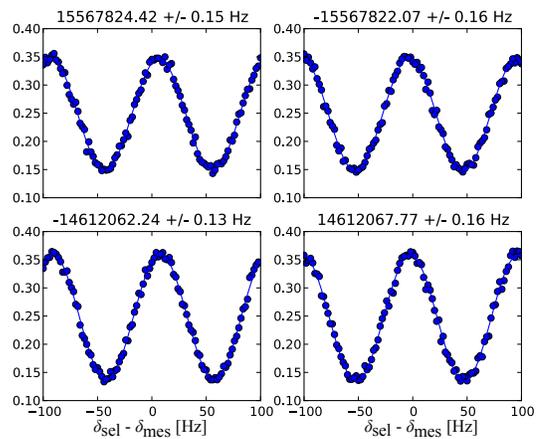}
\end{center}
\caption{Example of the four spectra needed to deduce $h/m_{\mathrm{Rb}}$ (see the text). Each spectrum represents the quantity $N_2/(N_1+N_2)$ where $N_1$ and $N_2$ are the population of the $F$ = 1 and 2 levels in function of the frequency difference between the two pairs of $\pi/2$ pulses. The measured position of the central fringe is indicated above the spectra.}
\label{4spectres}
\end{figure}

A value of $h/m_{\mathrm{Rb}}$ is obtained by recording four spectra. To cancel the velocity variation $gT$ due to the gravity $g$ ($T$ = 19 ms is the delay between the two pairs of $\pi/2$ pulses), the atoms are accelerated alternatively upward and downward and the difference between the results eliminate $gT$. Moreover, for each initial acceleration, two spectra are recorded by exchanging the directions of the Raman beams to eliminate the parasitic level shifts due to the Zeeman effect or to the light shifts. The figure~\ref{4spectres} shows an example of records. The acquisition time is about five minutes. For each spectrum, the Doppler shift is obtained with a relative accuracy of about $10^{-8}$. The ratio $\hbar/m_{\mathrm{Rb}}$ can be then determined from:
\begin{equation}
\frac{\hbar}{m_{\mathrm{Rb}}}= \frac{1}{4} \sum_\mathrm{4\ spectra} \frac{2\pi|\delta_{sel}-\delta_{mes}|}{2Nk_B(k_1+k_2)}
\label{hsurm}
\end{equation}
where $N$ = 500 is the number of Bloch oscillations in both opposite directions, $k_B$ is the Bloch wavevector and $k_1$ and $k_2$ are the wavevectors of the two Raman beams. Consequently, the spectra of Fig. \ref{4spectres} give $h/m_{\mathrm{Rb}}$ with a relative statistical uncertainty of $5 \times 10^{-9}$ ($2.5 \times 10^{-9}$ for $\alpha$).
\begin{figure}
\begin{center}
\includegraphics[width = .8\linewidth]{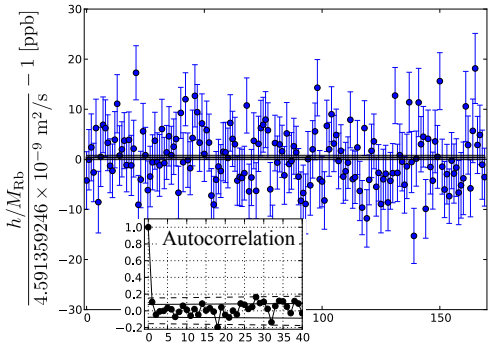}
\end{center}
\caption{Measurements of the ratio $h/m_{\mathrm{Rb}}$ during about 15 hours. The standard deviation of the mean is $4.4\times 10^{-10}$ with $\chi^2/(n-1)$ = 1.05. The inset shows the autocorrelation function of these 170 measurements. The solid and dashed lines represent the 1 $\sigma$ and 2 $\sigma$ standard deviation of the autocorrelation function.}
\label{pointshsurm}
\end{figure}
We have recorded about 1370 spectra analog to the figure~\ref{4spectres}. An example of 170 determinations of $h/m_{\mathrm{Rb}}$ obtained during one night is displayed on Fig. \ref{pointshsurm}. The autocorrelation function of these measurements (see the reference \cite{Witt}), which is reported in the inset of Fig. \ref{pointshsurm}, shows no correlation between the successive measurements of $h/m_{\mathrm{Rb}}$.

\begin{figure}
\begin{center}
\includegraphics[width = .8\linewidth]{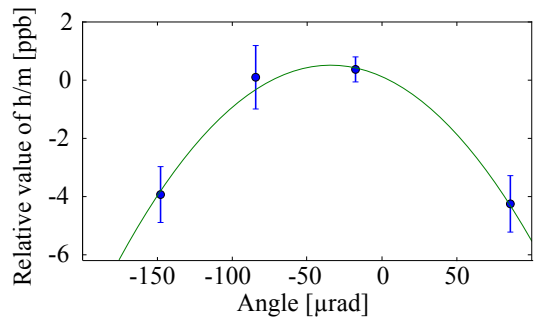}
\end{center}
\caption{(Color online). Sensitivity of the $h/m_{\mathrm{Rb}}$ measurements to the alignment: results of the measurements when the angle between the two Bloch beams is modified. The most precise point corresponds to the measurements in Fig. \ref{pointshsurm}. The difference between this value and the summit of the parabola which is fitted to the data is 0.35 ppb.}
\label{parabole}
\end{figure}

Table \ref{BudgetError} gives the error budget. The systematic effects are reduced compared to our previous measurements \cite{{Clade2},{Cadoret}}. The lasers are locked on a Fabry-Perot cavity stabilized with a standard laser and their frequencies are frequently measured with a frequency comb to reduce the frequency uncertainties at less than 50 kHz. The maximum angle between the lasers used for the Raman transitions and the Bloch oscillations is estimated to 40 $\mu$rad from the coupling between the two optical fibers. Moreover this value has been confirmed by the observation of the effect of the misalignment between the Bloch beams (see Fig. \ref{parabole}). The effect of the Gouy phase and the wave front curvature, which has been reduced by increasing the waist of the laser beam from $w$ = 2 mm to $w$ = 3.6 mm, has been carefully controlled with a Shack-Hartmann wave front analyzer. The parasitic magnetic field has been reduced with a double magnetic shield and a precise mapping of the magnetic field gives now a relative correction of $4\times 10^{-10}$. Thanks to the good collimation of the laser beams, the section of the laser beams varies by about $4\times 10^{-3}$ along the atomic trajectory and the result is a very good cancelation of the light shift effects between the upward and downward trajectory. From the density of the cloud of cold atoms after the RF selection and the two first sequences of BO (about $2\times 10^8$ atoms/cm$^3$), the effects of the refractive index and of the interactions between the atoms are estimated at a $10^{-10}$ level, corresponding to a conservative uncertainty of $2\times 10^{-10}$ in Table \ref{BudgetError}.  Thanks to the double cell with a differential pumping, the effect of the refractive index due to the background vapor (about $10^7$ atoms/cm$^3$) is now at the negligible level of few $10^{-11}$.
\begin{table}
\caption{\label{BudgetError} Error budget on the determination of
$1/ \alpha$ (systematic effect and relative uncertainty in part per $10^{10}$. }
\begin{ruledtabular}
\begin{tabular}{lcc}
\multicolumn{1}{l}{Source}
&Correction  &\parbox{2cm}{Relative uncertainty }\\
\hline Laser frequencies& &1.3\\
Beams alignment&-3.3& 3.3\\
Wavefront curvature and Gouy phase&-25.1 & 3.0\\
2nd order Zeeman effect&4.0 & 3.0 \\
Gravity gradient&-2.0& $0.2$ \\
Light shift (one photon transition)& & 0.1\\
Light shift (two photon transition)& & 0.01 \\
Light shift (Bloch oscillation)& & 0.5 \\
Index of refraction atomic cloud & & \\
and atom interactions& &2.0 \\ \hline
Global systematic effects&-26.4 & 5.9\\ \hline 
Statistical uncertainty& & 2.0\\
Rydberg constant and mass ratio & &2.2 \\ \hline \hline
Total uncertainty& & 6.6\\
\end{tabular}
\end{ruledtabular}
\end{table}

Taking into account all these corrections, the measured value of $h/m_{\mathrm{Rb}}$ is $4.591~359~2729~(57) \times 10^{-9}$m$^2$s$^{-1}$, and, using the values of $R_\infty$, $m_{\mathrm{e}}$ \cite{codata06} and $m_{\mathrm{Rb}}$ \cite{Myers}, we obtain the value of $\alpha$ given by equation (\ref{valeuralpha}).

\begin{figure}
\begin{center}
\includegraphics[width = .75\linewidth]{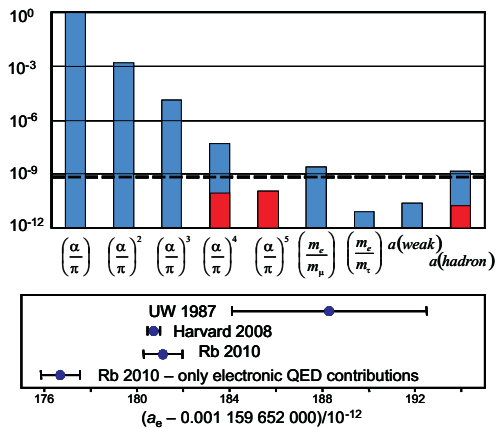}
\end{center}
\caption{(Color online). Upper figure: in blue, relative contributions to the electron anomaly of the different terms of equation (\ref{eqAnomalie}), in red their uncertainties. The dashed line corresponds to the relative uncertainty of the new $\alpha$ value. Lower figure: comparison of the measurements of the electron anomaly (UW 1987 \cite{VanDick} and Harvard 2008 \cite{Gabrielse2008}) with the theoretical value obtained by using the new value of $\alpha$ (label Rb 2010). The point "Rb 2010 - only QED" is obtained without the last term of equation (\ref{eqAnomalie}).}
\label{anomalie}
\end{figure}

With this new result the QED can be tested at a level better than $10^{-9}$. During two decades, the theory of $a_{\mathrm{e}}$ has been improved by Kinoshita and collaborators \cite{{Kinoshita},{Kinoshita1},{Kinoshita2}}. The anomaly is expressed as a sum of terms in power of $\alpha/\pi$ and of additive terms which take into account the contributions due to the muon, the tau, the weak interaction and the hadrons:
\begin{eqnarray}
a_\mathrm{e}&=&A_1\frac{\alpha}{\pi}+A_2\left(\frac{\alpha}{\pi}\right)^2+A_3\left(\frac{\alpha}{\pi}\right)^3
+A_4\left(\frac{\alpha}{\pi}\right)^4+ ... \nonumber \\
&&+a\left(\frac{m_\mathrm{e}}{m_\mu},\frac{m_\mathrm{e}}{m_\tau},\mathrm{weak},\mathrm{hadron}\right).
\label{eqAnomalie}
\end{eqnarray}
This equation and the new value of $\alpha$ give a theoretical value of the anomaly
$a_\mathrm{e}=0.001~159~652~181~13(84)$ where the uncertainty is due to the theory ($33\times 10^{-14}$) and to the $\alpha$ measurement ($78\times 10^{-14}$). The comparison with the experimental result \cite{Gabrielse2008} gives the difference $a_\mathrm{e}(\mathrm{theory})-a_\mathrm{e}(\mathrm{exp}.)=(40\pm89)\times10^{-14}$. The relative agreement between the experiment and the theory is at the level of $7.7\times10^{-10}$. The figure \ref{anomalie} shows the contributions of the different terms of equation (\ref{eqAnomalie}) (upper part) and the comparison between the experimental and theoretical values (lower part). The accuracy of the $\alpha$ measurement is sufficient to test for the first time the contributions due to the muon and hadrons. If we suppose the exactness of the QED calculation, this very good agreement provides a strong limitation to a possible structure of the electron \cite{Kinoshita2} or to the existence of new dark matter particles \cite{Boehm}.

In conclusion we have presented a recoil-velocity measurement of Rubidium and obtained a new determination of the fine structure constant with a relative uncertainty of $6.6\times 10^{-10}$. The combination of this result with the measurement and the calculation of the electron anomaly provides the most stringent test of QED. In the future, the sensibility of the interferometer can be increased by using a larger area interferometer \cite{{Muller},{LargeMomentum}} and the experiment can be improved to reduce the correction due to the Gouy phase and the wave front curvature and divide the uncertainty by a factor of two. Then a main limitation will be the uncertainty of the mass ratio $m_\mathrm{Rb}/m_\mathrm{e}$.

This experiment is supported in part by IFRAF (Institut Francilien de Recherches sur les Atomes Froids), and by the Agence Nationale pour la Recherche, FISCOM Project-(ANR-06-BLAN-0192).


\end{document}